\documentclass[twocolumn,showpacs,prb]{revtex4}

\usepackage{dcolumn}
\usepackage{bm}
\usepackage{graphicx}

\begin{document}

\title{Resonance Damping in Ferromagnets and Ferroelectrics}

\author{A. Widom}
\affiliation{Physics Department, Northeastern University, Boston, MA USA}

\author{S. Sivasubramanian}
\affiliation{NSF Nanoscale Science \& Engineering Center for High-rate Nanomanufacturing, 
\\ Northeastern University, Boston MA USA}

\author{C. Vittoria and S. Yoon}
\affiliation{Department of Electrical and Computer Engineering, 
Northeastern University, Boston, MA USA}

\author{Y.N. Srivastava}
\affiliation{Physics Department and \& INFN, University of Perugia, Perugia IT}

\begin{abstract}

The phenomenological equations of motion for the relaxation of ordered 
phases of magnetized and polarized crystal phases can be developed in close 
analogy with one another. For the case of magnetized systems, the driving 
magnetic field intensity toward relaxation was developed by Gilbert. For the 
case of polarized systems, the driving electric field intensity toward 
relaxation was developed by Khalatnikov. The transport times for relaxation 
into thermal equilibrium can be attributed to viscous sound wave damping via 
magnetostriction for the magnetic case and electrostriction for the 
polarization case.  

\end{abstract}


\maketitle

\section{Introduction \label{intro}}

It has long been of interest to understand the close analogies between ordered 
electric polarized systems, e.g. {\em ferroelectricity}, and ordered magnetic 
systems, e.g. {\em ferromagnetism}. At the microscopic level, the source of such 
ordering must depend on the nature of the electronic energy spectra. The relaxation 
mechanism into thermal equilibrium state must be described by local electric field 
fluctuations for the electric polarization case and by magnetic intensity 
fluctuations for the magnetization case; Specifically, the field fluctuations for 
each case   
\begin{eqnarray}
{\cal G}^{pol}_{ij}({\bf r},{\bf r}^\prime ,t)=\frac{1}{\hbar }\int_0^\beta 
\left<\Delta E_j({\bf r}^\prime ,-i\lambda)
\Delta E_i({\bf r} ,t)\right>d\lambda , 
\nonumber \\ 
{\cal G}^{mag}_{ij}({\bf r},{\bf r}^\prime ,t)=\frac{1}{\hbar }\int_0^\beta 
\left<\Delta H_j({\bf r}^\prime ,-i\lambda)
\Delta H_i({\bf r} ,t)\right>d\lambda ,
\nonumber \\ 
{\rm wherein} \ \ \beta=\frac{\hbar }{k_B T}\ ,
\label{intro1}
\end{eqnarray}   
determine the relaxation time tensor for both cases via the fluctuation-dissipation 
formula\cite{Machlup:1953I,Machlup:1953II,Callen:1962,Kubo:1998II}
\begin{equation}
\tau_{ij}=\int_0^\infty \lim_{V\to \infty}\left[\frac{1}{V} \int_V \int_V
{\cal G}_{ij}({\bf r},{\bf r}^\prime ,t)d^3{\bf r} d^3{\bf r}^\prime \right]dt .
\label{intro2}
\end{equation} 
We have {\em unified the theories of relaxation} in ordered polarized 
systems and ordered magnetized systems via the Kubo transport time tensor in 
Eqs.(\ref{intro1}) and (\ref{intro2}).  

The transport describing the relaxation of ordered magnetization is 
the Landau-Lifshitz-Gilbert equation\cite{Landau:1935,Gilbert:1955,Gilbert:2004}. 
This equation has been of considerable recent 
interest\cite{Fahnle:2008,Rossi:2005,Kambersky:2007} 
in describing ordered magnetic resonance 
phenomena\cite{Garate:2009,Brataas:2008,Gilmore:2007,McMichael:2007}. 
The equation describing the electric relaxation of an ordered polarization is 
the Landau-Khalatinikov-Tani 
equation\cite{Landau:1954,Makita:1970,Tani:1968}. This equation can be simply 
modeled\cite{Sivasubramanian:2004,Li:2008,In:2008,Gordon:2008} 
with effective electrical 
circuits\cite{Sivasubramanian:2003,Li:2009,Guyomar:2007,Ando:2009}. 
Information memory 
applications\cite{Yoshihisa:2007,Kohlstedt:2005,Sun:2008,Jones:1998} 
of such polarized system are of considerable recent 
interest\cite{Vrejoiu:2009,Inoue:2001,Mikolajick:2001}. 

The unification of the magnetic Gilbert-Landau-Lifshitz equations   
and the electric Landau-Khalatnikov-Tani equations via the 
relaxation time tensor depends on the notion of a {\em nonequilibrium driving field}. 
For the magnetic case, the driving magnetic intensity 
\begin{math} {\bf H}_d \end{math} determines the relaxation of the magnetization 
via the torque equation 
\begin{equation}
\dot{\bf M}=\gamma {\bf M}\times {\bf H}_d \ ,
\label{intro3}
\end{equation} 
wherein \begin{math} \gamma \end{math} is the gyromagnetic ratio.  
For the electric case, the driving electric field 
\begin{math} {\bf E}_d \end{math} determines the relaxation of the polarization  
via the equation of motion for an ion of charge \begin{math} ze \end{math} 
\begin{equation}
m\ddot{\bf r}=ze{\bf E}_d .
\label{intro4}
\end{equation} 
The unification of both forms of relaxation lies in the close analogy between 
the magnetic driving intensity \begin{math} {\bf H}_d \end{math} and the electric 
driving field \begin{math} {\bf E}_d \end{math}.  

In Sec.\ref{thermo} the thermodynamics of ordered magnetized and polarized 
systems is reviewed. The notions of magnetostriction and electrostriction 
are given a precise thermodynamic definition. In Sec.\ref{res}, the 
phenomenology of the relaxation equations are presented. The magnetic driving 
intensity \begin{math} {\bf H}_d \end{math} and the electric driving field 
\begin{math} {\bf E}_d \end{math} are defined in terms of the relaxation time 
tensor Eq.(\ref{intro2}). In Sec.\ref{sh}, we introduce the crystal viscosity 
tensor. From a Kubo formula viewpoint, the stress fluctuation correlaxation  
\begin{equation}
{\cal F}_{ijkl}({\bf r},{\bf r}^\prime ,t)=\frac{1}{\hbar }\int_0^\beta 
\left<\Delta \sigma_{kl}({\bf r}^\prime ,-i\lambda)
\Delta  \sigma_{ij}({\bf r} ,t)\right>d\lambda , 
\label{intro5}
\end{equation}   
determines the crystal viscosity  
\begin{eqnarray}
\eta_{ijkl}= 
\nonumber \\ 
\int_0^\infty \lim_{V\to \infty}\left[\frac{1}{V} \int_V \int_V
{\cal F}_{ijkl}({\bf r},{\bf r}^\prime ,t)d^3{\bf r} d^3{\bf r}^\prime \right]dt .
\label{intro6}
\end{eqnarray} 
For models of magnetic relaxation wherein acoustic heating dominates via  
magnetostriction\cite{Vittoria:2010} 
and for models of electric relaxation wherein acoustic heating 
dominates via electrostriction, the relaxation time tensor in Eq.(\ref{intro2}) 
can be related to the viscosity tensor Eq.(\ref{intro6}). An independent microscopic 
derivation of viscosity induced relaxation is given in Appendix \ref{fd}.
In the concluding Sec.\ref{conc}, the sound wave absorption physics of the viscous 
damping mechanism will be noted.  

\section{Thermodynamics \label{thermo}}

Our purpose is to review the thermodynamic properties of both magnetically 
ordered crystals and polarization ordered crystals. The former is characterized 
by a remnant magnetization \begin{math} {\bf M} \end{math} for vanishing 
applied magnetic intensity \begin{math} {\bf H}\to 0\end{math} while the latter 
is characterized by a remnant polarization \begin{math} {\bf P} \end{math} 
for vanishing applied electric field \begin{math} {\bf E}\to 0\end{math}.

\subsection{Magnetically Ordered Crystals \label{moc}} 

Let \begin{math} w \end{math} be the enthalpy per unit volume. The fundamental 
thermodynamic law determining the equations of state for magnetically ordered 
crystals is given by 
\begin{equation}
dw=Tds+{\bf H}\cdot d{\bf M}-{\bm e}:d{\bm \sigma},
\label{moc1}
\end{equation} 
wherein \begin{math} s \end{math} is the entropy per unit volume, 
\begin{math} T \end{math} is the temperature, \begin{math} {\bm e} \end{math} 
is the crystal strain and \begin{math} {\bm \sigma } \end{math} is the crystal 
stress. The magnetic adiabatic susceptibility is defined by 
\begin{equation}
{\bm \chi}=
\left(\frac{\partial {\bf M}}{\partial {\bf H}}\right)_{s,{\bm \sigma }}.
\label{mocchi}
\end{equation}
If 
\begin{equation} 
{\bf N}=\frac{{\bf M}}{M}\ \ \ \ \Rightarrow \ \ \ \ {\bf N\cdot N}=1
\label{mocunit} 
\end{equation} 
denotes a unit vector in the direction of the magnetization, then the tensor 
\begin{math} \Lambda_{ijkl} \end{math} describing adiabatic magnetostriction 
coefficients may be defined as\cite{Landau:1960} 
\begin{equation}
2\Lambda_{ijkl}N_l=
M\left(\frac{\partial e_{ij}}{\partial M_k}\right)_{s,{\bm \sigma}}=
-M\left(\frac{\partial H_k}{\partial \sigma_{ij}}\right)_{s,{\bf M}}.
\label{moc2}
\end{equation}
When the system is out of thermal equilibrium, the driving magnetic intensity is 
\begin{equation}
{\bf H}_d={\bf H}-\left(\frac{\partial w}{\partial {\bf M}}\right)_{s,{\bm \sigma}} 
-{\bm \tau}\cdot \left(\frac{\partial {\bf M}}{\partial t}\right),
\label{moc3}
\end{equation}
wherein \begin{math} {\bm \tau}  \end{math} are the relaxation time tensor 
transport coefficients which determine the relaxation of the ordered magnetic system 
into a state of thermal equilibrium. 

\subsection{Ordered Polarized Crystals \label{opc}} 

The fundamental thermodynamic law determining the equations of state for ordered 
polarized crystals is given by 
\begin{equation}
dw=Tds+{\bf E}\cdot d{\bf P}-{\bm e}:d{\bm \sigma},
\label{opc1}
\end{equation} 
wherein \begin{math} w \end{math} is the enthalpy per unit volume, 
\begin{math} s \end{math} is the entropy per unit volume, 
\begin{math} T \end{math} is the temperature, \begin{math} {\bm e} \end{math} 
is the crystal strain and \begin{math} {\bm \sigma } \end{math} is the crystal 
stress. 
The electric adiabatic susceptibility is defined by 
\begin{equation}
{\bm \chi}=
\left(\frac{\partial {\bf P}}{\partial {\bf E}}\right)_{s,{\bm \sigma }}.
\label{opcchi}
\end{equation}
The tensor \begin{math} \beta_{ijk} \end{math} describing adiabatic 
electrostriction coefficients may be defined as\cite{Landau:1960}
\begin{equation}
\beta_{ijk}=
\left(\frac{\partial e_{ij}}{\partial P_k}\right)_{s,{\bm \sigma}}=
-\left(\frac{\partial E_k}{\partial \sigma_{ij}}\right)_{s,{\bf P}}.
\label{opc2}
\end{equation}
The piezoelectric tensor is closely related to the electrostriction tensor via 
\begin{equation}
\gamma_{ijk}=
\left(\frac{\partial e_{ij}}{\partial E_k}\right)_{s,{\bm \sigma}}=
\left(\frac{\partial P_k}{\partial \sigma_{ij}}\right)_{s,{\bf E}}
=\beta_{ijm}\chi_{mk}\ .
\label{opcgamma}
\end{equation}
When the system is out of thermal equilibrium, the driving electric field is  
\begin{equation}
{\bf E}_d={\bf E}-\left(\frac{\partial w}{\partial {\bf P}}\right)_{s,{\bm \sigma}} 
-{\bm \tau}\cdot \left(\frac{\partial {\bf P}}{\partial t}\right),
\label{opc3}
\end{equation}
wherein \begin{math} {\bm \tau}  \end{math} is the relaxation time tensor 
transport coefficients which determine the relaxation of the ordered polarized 
system into a state of thermal equilibrium. 

\section{Resonance Dynamics \label{res}}

Here we shall show how the magnetic intensity 
\begin{math} {\bf H}_d \end{math} drives the magnetic resonance equations of 
motion in magnetically ordered systems. Similarly, we shall show how the electric field 
\begin{math} {\bf E}_d \end{math} drives the polarization resonance equations of 
motion for polarized ordered systems.

\subsection{Gilbert-Landau-Lifshitz Equations \label{ge}} 

The driving magnetic intensity determines the torque on the magnetic 
moments according to 
\begin{equation}
\frac{\partial {\bf M}}{\partial t}=\gamma {\bf M}\times {\bf H}_d\ .
\label{ge1}
\end{equation}
Employing 
Eqs.(\ref{moc3}) and (\ref{ge1}), one finds the equations for magnetic 
resonance in the Gilbert form 
\begin{equation}
\frac{\partial {\bf M}}{\partial t}=\gamma {\bf M}\times 
\left[{\bf H}-\left(\frac{\partial w}{\partial {\bf M}}\right)_{s,{\bm \sigma}} 
-\left(\frac{{\bm \alpha}}{\gamma M}\right)\cdot 
\frac{\partial {\bf M}}{\partial t} \right],
\label{ge2}
\end{equation}
wherein the Gilbert dimensionless damping tensor 
\begin{math} {\bm \alpha} \end{math} is defined as  
\begin{equation}
{\bm \alpha }=(\gamma M){\bm \tau }.
\label{ge3}
\end{equation}
One may directly solve the Gilbert equations for the driving magnetic intensity 
according to 
\begin{equation}
{\bf H}_d +{\bf \alpha }\cdot \big({\bf N}\times {\bf H}_d\big)=
{\bf H}-\left(\frac{\partial w}{\partial {\bf M}}\right)_{s,{\bm \sigma}}.
\label{ge4}
\end{equation}
Eqs.(\ref{ge1}) and (\ref{ge4}) express the magnetic resonance motion in the 
Landau-Lifshitz form.

\subsection{Landau-Khalatnikov-Tani Equations \label{ke}} 

The driving electric field gives rise to a polarization response according to 
\begin{equation}
\frac{\partial ^2 {\bf P}}{\partial t^2}=
\left(\frac{\omega_p^2}{4\pi }\right){\bf E}_d ,
\label{ke1}
\end{equation}
wherein \begin{math} \omega_p \end{math} is the plasma frequency. A simple derivation 
of Eq.(\ref{ke1}) may be formulated as follows. In a large volume 
\begin{math} V \end{math}, the polarization due to charges 
\begin{math} \{z_je \} \end{math} is given by 
\begin{equation}
{\bf P}=\left(\frac{\sum_j z_je {\bf r}_j}{V}\right).
\label{ke2}
\end{equation}
If the driving electric field accelerates the charges according to 
\begin{equation}
m_j \ddot{\bf r}_j=z_j e {\bf E}_d,
\label{ke3}
\end{equation}
then Eq.(\ref{ke1}) holds true with the plasma frequency
\begin{equation}
\omega_p^2=
4\pi e^2\lim_{V\to \infty}\left[\frac{\sum_j (z_j^2/m_j)}{V}\right]
=4\pi e^2\sum_a \frac{n_a z_a^2}{m_a}\ ,
\label{ke4}
\end{equation}
wherein \begin{math} n_a \end{math} is the density of charged particles of type {\it a}.

The polarization resonance equation of motion follows from Eqs.(\ref{opc3}) and 
(\ref{ke1}) as\cite{Tani:1968}  
\begin{equation}
\left(\frac{4\pi }{\omega_p^2}\right)\frac{\partial^2 {\bf P}}{\partial t^2}+
{\bm \tau }\cdot \frac{\partial {\bf P}}{\partial t}+
\frac{\partial w({\bf P},s,{\bm \sigma})}{\partial {\bf P}}={\bf E}.
\label{ke5}
\end{equation}
The electric field \begin{math} {\bf E} \end{math} induces the polarization  
\begin{math} {\bf P} \end{math} at resonant frequencies which are eigenvalues of 
the tensor \begin{math} {\bm \Omega} \end{math} for which 
\begin{equation}
{\bm \Omega}^2 = \frac{\omega_p^2 {\bm \chi }^{-1}}{4\pi }
\equiv \omega_p^2 ({\bm \epsilon}-{\bf 1})^{-1} \ .
\label{ke6}
\end{equation}
The decay rates for the polarization oscillations are eigenvalues of 
the tensor \begin{math} {\bm \Gamma } \end{math} for which 
\begin{equation}
{\bm \Gamma }=\frac{\omega_p^2 {\bm \tau }}{4\pi }\ .
\label{ke7}
\end{equation}
If the decay rates are large on the scale of the the resonant frequencies, 
then the equation of motion is over damped so that 
\begin{eqnarray}
\min_j\Gamma_j \gg \max_i \Omega_i \ \ \ {\rm implies}\ 
\nonumber \\ 
{\bm \tau }\cdot \frac{\partial {\bf P}}{\partial t}+
\frac{\partial w({\bf P},s,{\bm \sigma})}{\partial {\bf P}}={\bf E}.
\label{ke8}
\end{eqnarray} 
Eq.(\ref{ke8}) represents the Landau-Khalatnikov equation for polarized 
systems.

\section{Heating Rate per Unit Volume\label{sh}}

Let us here consider the heating rate implicit in relaxation processes.
Independently of the details of the microscopic mechanism for generating 
such heat, the rates of energy dissipation are {\em entirely determined} by 
\begin{math} {\bm \tau}  \end{math}.
Explicitly, the heating rates per unit volume for magnetization and 
polarization are given, respectively, by 
\begin{equation}
\dot{q}_{\bf M}=\frac{\partial {\bf M}}{\partial t}\cdot {\bm \tau}
\cdot \frac{\partial {\bf M}}{\partial t}\ ,
\label{sh1a}
\end{equation}
and 
\begin{equation}
\dot{q}_{\bf P}=\frac{\partial {\bf P}}{\partial t}\cdot {\bm \tau}
\cdot \frac{\partial {\bf P}}{\partial t}\ .
\label{sh1b}
\end{equation}
Finally, the notion of crystal viscosity \begin{math} \eta_{ijkl} \end{math} 
is introduced into elasticity theory\cite{Landau:1986} via the heating rate 
per unit volume from rates of change in the strain 
\begin{math} \partial {\bm e}/\partial t \end{math}; It is  
\begin{equation}
\dot{q}_{\bm e}=\frac{\partial e_{ij}}{\partial t}
\ \eta_{ijkl}\ \frac{\partial e_{kl}}{\partial t}.
\label{sh2}
\end{equation}
Crystal viscosity is employed to describe, among other things,   
sound wave attenuation. Our purpose is to describe how heating rates in 
Eqs.(\ref{sh1a}) and (\ref{sh1b}) can be related to the heating rate 
in Eq.((\ref{sh2})). This allows us to express the transport coefficients 
\begin{math} {\bm \tau} \end{math} in terms of the crystal viscosity. 

\subsection{Relaxation via Magnetostriction \label{rm}}

From the magnetostriction Eq.(\ref{moc2}), it follows that magnetic 
relaxation gives rise to a strain  
\begin{equation}
\frac{\partial e_{ij}}{\partial t}=
\frac{2}{M}\Lambda_{ijkl}N_k\frac{\partial M_l}{\partial t}, 
\label{rm1}
\end{equation} 
and thereby to the heating rate, 
\begin{equation}
\dot{q}=\frac{4}{M^2}\frac{\partial M_i}{\partial t}
(\Lambda_{mnqi}N_q) \eta_{mnrs} (\Lambda_{rskj}N_k)
\frac{\partial M_j}{\partial t}\ ,
\label{rm2}
\end{equation}
in virtue of Eq.(\ref{sh2}). Employing Eqs.(\ref{sh1a}) and (\ref{rm2}), 
we find that the magnetic relaxation transport coefficient in the magnetostriction  
model 
\begin{equation}
\tau_{ij}=\frac{4}{M^2}(\Lambda_{mnqi}N_q )\eta_{mnrs} (\Lambda_{rskj}N_k) .
\label{rm3}
\end{equation}
The Gilbert damping tensor follows from Eqs.(\ref{ge3}) and (\ref{rm3}) as 
\begin{equation}
\alpha_{ij}=\frac{4\gamma }{M}(\Lambda_{mnqi}N_q )\eta_{mnrs} (\Lambda_{rskj}N_k) .
\label{rm4}
\end{equation}
The central relaxation tensor Eq.(\ref{rm4}) describes the magnetic relaxation in 
terms of the magnetostriction coefficients and the crystal viscosity. 

\subsection{Relaxation via Electrostriction \label{re}}

From the electrostriction Eq.(\ref{opc2}), it follows that 
a time varying polarization gives rise to a time varying strain  
\begin{equation}
\frac{\partial e_{ij}}{\partial t}=
\beta_{ijk}\frac{\partial P_k}{\partial t}, 
\label{re1}
\end{equation} 
and thereby to the heating rate, 
\begin{equation}
\dot{q}=\frac{\partial P_i}{\partial t}
\beta_{kli}\eta_{klmn}\beta_{mnj}\frac{\partial P_j}{\partial t}\ ,
\label{re2}
\end{equation}
in virtue of Eq.(\ref{sh2}). Employing Eqs.(\ref{sh1b}) and (\ref{re2}), 
we find that the electric relaxation transport coefficient in the electroostriction  
model 
\begin{equation}
\tau_{ij}=\beta_{kli} \eta_{klmn} \beta_{mnj} .
\label{re3}
\end{equation}
The central relaxation tensor Eq.(\ref{re3}) describes the polarization relaxation time  
tensor coefficients in terms of the electrostriction coefficients and the 
crystal viscosity. The implications of the electrostriction model for the 
Landau-Khalatnikov equation is to the authors knowledge a new result.

\section{Conclusions \label{conc}}

For ordered polarized and magnetized systems, we have developed 
phenomenological equations of motion in close analogy with one another.
For the magnetized case, the relaxation is driven by the magnetic intensity 
\begin{math} {\bf H}_d \end{math} yielding the Gilbert  
equation of motion\cite{Gilbert:2004}. For the polarized case, the relaxation 
is driven by the electric field  \begin{math} {\bf E}_d \end{math} yielding 
the Tani equation of motion\cite{Tani:1968}. In both cases, the relaxation 
time tensor \begin{math} {\bm \tau } \end{math} is determined by the crystal 
viscosity as derived in the Appendix \ref{fd}; i.e. in Eqs.(\ref{fd3}) and 
(\ref{fd6}). The viscosity can be measured independently from the magnetic or 
electrical relaxation by employing sound absorption 
techniques\cite{Bhatia:1967}.

\medskip
\appendix 
\section{Kubo formulae \label{fd}}

From the thermodynamic Eq.(\ref{moc2}), the fluctuations in the magnetic 
intensity are given by magnetostriction, i.e. 
\begin{equation}
\Delta H_k({\bf r},t)=-\left(\frac{2\Lambda_{ijkl}N_l}{M}\right)
\Delta \sigma_{ij}({\bf r},t).  
\label{fd1}
\end{equation}
Eqs.(\ref{fd1}), (\ref{intro1}) and (\ref{intro5}) imply 
\begin{eqnarray}
{\cal G}^{mag}_{ij}({\bf r},{\bf r}^\prime ,t)= 
\nonumber \\ 
\frac{4}{M^2}(\Lambda_{mnqi}N_q )
{\cal F}_{mnrs}({\bf r},{\bf r}^\prime ,t)(\Lambda_{rskj}N_k) .
\label{fd2}
\end{eqnarray}
Employing Eqs.(\ref{fd2}), (\ref{intro2}) and (\ref{intro6}), one finds the 
central result for the magnetic relaxation time tensor; It is  
\begin{equation}
\tau^{mag}_{ij}=
\frac{4}{M^2}(\Lambda_{mnqi}N_q )\eta_{mnrs} (\Lambda_{rskj}N_k)
=\frac{\alpha_{ij}}{\gamma M}\ .
\label{fd3}
\end{equation}

From the thermodynamic Eq.(\ref{opc2}), the fluctuations in the electric  
intensity are given by electrostriction, i.e.  
\begin{equation}
\Delta E_k({\bf r},t)=-\beta_{ijk}\Delta \sigma_{ij}({\bf r},t).  
\label{fd4}
\end{equation}
Eqs.(\ref{fd4}), (\ref{intro1}) and (\ref{intro5}) imply 
\begin{eqnarray}
{\cal G}^{pol}_{ij}({\bf r},{\bf r}^\prime ,t)= 
\beta_{kli} {\cal F}_{klmn}({\bf r},{\bf r}^\prime ,t)\beta_{mnj}. 
\label{fd5}
\end{eqnarray}
Employing Eqs.(\ref{fd5}), (\ref{intro2}) and (\ref{intro6}), one finds the 
central result for the electric relaxation time tensor; It is  
\begin{equation}
\tau^{pol}_{ij}=\beta_{kli} \eta_{klmn} \beta_{mnj} .
\label{fd6}
\end{equation}

\end{document}